\documentstyle[aps,preprint]{revtex}

\begin{document}
\draft
\title{CHAINING IN MAGNETIC COLLOIDS IN THE PRESENCE OF FLOW }

\author{I. P\'erez-Castillo, A. P\'erez-Madrid, J. M. Rub\'{\i}}
\address{Departament de F\'{\i}sica Fonamental and CER F\'{\i}sica de Sistemes Complexos,\\
Facultat de F\'{\i}sica,\\
Universitat de Barcelona,\\
Diagonal 647, 08028 Barcelona, Spain\\}

\author{G. Bossis}
\address{Laboratoire de Physique de la Mati\`ere Condens\'ee,\\ CNRS UMR 6622, Parc Valrose,\\ 06108 Nice Cedex 2, France}
\maketitle

\maketitle
\begin{abstract}
We discuss the effect of an homogeneous flow in the aggregation process of colloidal magnetic particles at moderate concentration. Situations in which the presence of flow acts in favor of the chaining process: particles assemble into chains larger than the ones emerging in the absence of flow, under the only influence of an externally imposed field, have been analyzed. The results we obtain follow from the analysis of the pair correlation function which, owing to the potential character of the flow we consider, can be interpreted in terms of a Boltzmann-like stationary distribution function. To render the influence of the flow on the resulting structures explicit, we study the particular cases of axisymmetric and planar elongational flows.
\end{abstract}

\section{Introduction}
One of the most remarkable features of the many-particle regime of suspension of magnetic colloids is the formation of structures -mainly chains or branched structures- which may in general evolve over time$^{\ref{rubi_vilar}}$. The reason for the occurrence of this phenomenon is the presence of dipolar interactions among particles. In the absence of an external field, the form of the emerging structures depends basically on the ratio between dipolar and thermal energies. When dipolar energy is dominant, alignment  of particles is favored making the chain the most likely structure. The tendency to disorder, promoted by thermal motion, may destroy that alignment resulting in the appearance of more isotropic aggregates. These structures are in fact fractal objects and have been classified in terms of the fractal dimension$^{\ref{niklasson}, \ref{pastor_rubi}}$. A fractal dimension close to one indicates the presence of chain-like aggregates whereas higher values of that quantity  is the signature of DLA magnetic  aggregates$^{\ref{pastor_rubi2}}$.

When an external magnetic field is applied, the dipoles are oriented along the direction of the field thus favoring the formation of chain-like aggregates. This strongly anisotropic situation becomes manifest through the form of the pair correlation, revealing that the underlying structure is the chain$^{\ref{gennes}}$.

The earlier stages of the chaining process are characterized by the existence of a scaling regime for the size of the chain-like aggregates. In the absence of flow, the scaling behavior for the size of the chain $S\sim t^\alpha$ has been found with an exponent $\alpha = 1/2$, obtained under the approximation of spherical cluster $^{\ref{miyazima}}$. By considering the more realistic situation in which the diffusion process is mediated by anisotropic hydrodynamic interactions among particles, one obtains logarithmic corrections to the power laws providing the effective value of the exponent $\alpha_{eff} = 0.61$ $^{\ref{carmen}}$, which agrees with experimental results $^{\ref{fraden}}$. At longer times, one has to consider the breaking of the chains when they become larger than a characteristic length. This feature is responsible for the existence of a crossover to a stationary regime characterized by the average size of the chains. It is in  this regime in which we want to analyze the influence of the flow. 

The situations reported previously occur when aggregation phenomena take place in a quiescent liquid. In many cases of interest, i.e. porous media, active dampers, rheological devices to control the viscosity, or in general in rheology of field-responsive systems, aggregation proceeds under conditions creating a stationary flow. The presence of these chain-like structures in turns, affects the rheology of the system$^{\ref{clara} - \ref{carmen2}}$. The question is to what extent the flow constitutes an important factor to determine the shape of the emerging structures. This is precisely the problem we address in this paper. We study aggregation of the magnetic grains under an elongational flow. Owing to the potential nature of that flow, for which the stationary distribution function of a pair of particles is Boltzmann-like$^{\ref{clara}, \ref{kramers}}$, we can perform an analysis similar to the one carried out for aggregation in a quiescent liquid$^{\ref{gennes}}$. The knowledge of the pair correlation function or equivalently of the second virial coefficient will allow us to derive the equation of state in the presence of flow and to compute the mean size of a chain. 

We have organized the paper in the following way. In section 2 we describe the system and discuss the nature of the interactions between particles. Section 3 is devoted to calculate the pair correlation function and the related second virial coefficient corresponding to different situations concerning the relative orientation of the field and the external flow. The average size of a chain relative to the different situations discussed in the previous section is calculated in section 4. We extend the results of previous theories, expressing that quantity as a function of the magnetic energy between particles, to the case of an externally imposed elongational flow. Finally, in the last section we summarize the effects that the flow has on the kinetics of chaining processes.

\section{Interacting particles in external flow}

 We consider a colloidal suspension of magnetic dipoles in a non-polar liquid phase. The particles bear permanent magnetic moments rigidly attached. The one for the i-th particle is

\begin{equation}\label{eq:a1}
{\bf m}_i = m \hat{\bf R}_i\; ,
\end{equation}

\noindent where $\hat{\bf R}_i$ is the unit vector along the direction of ${\bf m}_i$  and $m$ is the constant magnetic moment strength. The system is under the influence of both a constant external magnetic field ${\bf H}$, high enough to induce alignment of the magnetic grains, and a stationary elongational flow, described by the velocity field 

\begin{equation}\label{eq:a2}
 v_i =  A_{ij} x_j\; ,
\end{equation}

\noindent where $A_{ij}$ are the components of the velocity gradient tensor, and $x_j$ is the j-th component of the position vector. 

The flow induces on a pair of bounded particles a stretching, with associated  energy $^{\ref{kramers}}$

\begin{equation}\label{eq:a5}
U_{fl} = -\frac{1}{2}\xi A_{ij}r_j r_i\, .
\end{equation}

\noindent which depends on the components of the vector ${\bf r}$ joining the center of the j-th and i-th spheres. Here $\xi$ is the Stokes friction coefficient. 

The potential energy of a pair of particles splits up into the following contributions 

\begin{equation}\label{eq:c1}
U = U_{fl}+ U_{mag}  
\end{equation}

\noindent where apart from the potential associated to the flow defined in Eq. (\ref{eq:a5})  we have introduced the energy $U_{mag}$ resulting from interactions between two magnetic dipoles a distance $r$ apart 

\begin{equation}\label{eq:a4}
U_{mag} = \frac{1}{r^3}\left\{{\bf m}_1\cdot{\bf m}_2-\frac{3}{r^2}({\bf m}_1\cdot{\bf r})({\bf m}_2\cdot{\bf r})\right\}\; .
\end{equation}

As follows from expressions (\ref{eq:a5})-(\ref{eq:a4}), the energy of a pair of particles, and consequently the kinetics of the chaining process, depends on both the velocity gradient, and the orientation of the magnetic field with respect to the symmetry axis of the flow. To analyze that dependence explicitly, two situations will be under scrutiny. First, we will consider a planar elongational flow, which can be generated by four rolling rolls placed at the corners of a square, for which the nonzero components of the velocity gradient tensor are given by $^{\ref{brenner}}$

\begin{equation}\label{eq:s2}
 A_{12} = A_{21} = A/2   \; ,
\end{equation}

\noindent and second an axisymmetric elongational flow, representing the flow through a pore, for which the nonzero components of the velocity gradient tensor are given by $^{\ref{brenner}}$
\begin{equation}\label{eq:s3}
A_{11} = A_{22} = -A/2  \; ; \; A_{33} = A\; ,
\end{equation}

\noindent with $A/2$ being the rate of elongation.

\section{The pair correlation function and the second virial coefficient}

For concentrations pertaining to the semi-dilute regime, the ferromagnetic particles tend to form chains along the direction of the imposed magnetic field. The system then reaches an equilibrium state in which the distribution function of the relative position of a pair of bounded particles is given by the corresponding Boltzmann distribution$^{\ref{gennes}}$. The flow acts as a perturbation such that the resulting stationary distribution is also a Boltzmann distribution in which the potential is given by Eq. (\ref{eq:c1}). It has been shown that this distribution is the stationary solution of the corresponding Fokker-Planck equation$^{\ref{clara}}$. 

The pair correlation function of the suspension is given by$^{\ref{goldstein}}$

\begin{equation}\label{eq:b2}
g(r) = \exp\left(-\frac{U}{k_BT}\right)
\end{equation}

\noindent where $T$ is the temperature and $k_B$ is the Boltzmann's constant. Its knowledge, allows us to derive the expression for the second virial coefficient, which is defined through the expression

\begin{equation}\label{eq:b4}
B(T) = -\frac{1}{2}\int_{\Omega} \left( g(r) -1\right) d^3{\bf r}\; .
\end{equation}

\noindent Here $\Omega$ represents the accessible volume for a particle of the chain.
 
In this section, the pair correlation and the second virial coefficient will be computed in the situations corresponding to the two types of flow introduced previously, taking into account different orientations of the magnetic field. For the planar elongational flow we will analyze the following cases:

\noindent a) The magnetic field is oriented in the direction forming an angle $\pi/4$ with respect to the x-axis (see inset of Fig. 1). In this case, the pair correlation function (\ref{eq:b2}) can be written as follows

\begin{equation}\label{eq:b3}
g(r) = \left\{\begin{array}{cl}
\exp\left\{\frac{P}{2}x^2 \sin^2\theta\sin(2\varphi)+ \lambda x^{-3}\left[\frac{3}{2}\sin^2\theta\left(1 + \sin(2\varphi)\right) -1 \right]\right\}&\mbox{; $x\geq 1$}\\
\\
0 &\mbox{; $x<1$}
\end{array}\right.
\end{equation}

\noindent where $x \equiv r/d$, with $d$ being the diameter of a dipole, and $0\leq\theta\leq\pi$ and $0\leq\varphi\leq 2\pi$ are the polar spherical angles. Moreover,  we have introduced the dimensionless parameters $\lambda = m^2/d^3k_BT$,which corresponds to the ratio between the dipolar and thermal energies, and the P\'eclet number, $P = \frac{\xi d^2 A/2}{2k_BT}$, expressing the ratio between hydrodynamic  and  thermal energies. Through this paper we will assume that $\lambda > P$. To arrive at this formula, we have taken into account the hard core interaction between two dipoles.  

By substitution of Eq. (\ref{eq:b3}) into Eq. (\ref{eq:b4}) it follows the asymptotic expression of $B(T)$ for large values of $\lambda$

\begin{equation}\label{eq:s4}
B = \frac{-\pi d^3 \exp(2\lambda+P/2)}{(6\lambda-P)\sqrt{(3\lambda+P)(3\lambda+P/2)}}\; .
\end{equation}

\noindent   In obtaining Eq. (\ref{eq:s4}) we have taken into account that the main contribution to the integral in Eq. (\ref{eq:b4}) corresponds to $r\approx d$, and to the region around the directions $\theta \approx \pi/2$, and $\varphi\approx \pi/4$ or $\varphi\approx 5\pi/4$. Notice that the dipoles link up preferably in the direction of the magnetic field, where the magnetic energy is attractive. Details about the derivation of Eq. (\ref{eq:s4}) have been given in the Appendix.

\noindent b) The magnetic field  is now oriented parallel to the x-axis (see inset in Fig. 2). The pair correlation function is given by

\begin{equation}\label{eq:s5}
g(r) = \left\{\begin{array}{cl}
\exp\left\{\frac{P}{2}x^2 \sin^2\theta\sin(2\varphi)+ \lambda x^{-3}(3\sin^2\theta\cos^2\varphi -1)\right\}&\mbox{; $x\geq 1$}\\
\\
0 &\mbox{; $x<1$}
\end{array}\right.
\end{equation}

\noindent In this case, the asymptotic  expression of $B(T)$ is

\begin{equation}\label{eq:s6}
B = -\frac{\pi d^3}{18\lambda^2} \exp(2\lambda+ P^2/12\lambda)\left(1 + \frac{1}{3\lambda}\right)\; .
\end{equation}

\noindent This result follows after consideration of the main contributions to the integral in Eq. (\ref{eq:b4}), which now correspond to  $r\approx d$, $\theta \approx \pi/2$, and $\varphi\approx 0$ or $\varphi\approx \pi$.

\noindent c) The magnetic field is perpendicular to the plane of the flow. In this case, for  the pair correlation function (\ref{eq:b2}) we obtain

\begin{equation}\label{eq:s7}
g(r) = \left\{\begin{array}{cl}
\exp\left\{\frac{P}{2}x^2 \sin^2\theta\sin(2\varphi)+ \lambda x^{-3}(3\cos^2\theta\cos^2\varphi -1)\right\}&\mbox{; $x\geq 1$}\\
\\
0 &\mbox{; $x<1$}
\end{array}\right.
\end{equation}

\noindent Analogously, the asymptotic expression of $B(T)$ is now given by

\begin{equation}\label{eq:s8}
B = -\frac{\pi d^3\exp(2\lambda)}{6\lambda\sqrt{9\lambda^2- P^2/4}}  \left(1 + \frac{1}{3\lambda}\right)\left(1- \frac{1}{2}\frac{\lambda}{9\lambda^2- P^2/4}\right)\; .
\end{equation}

\noindent In deriving this expression we have used the approximation that the  integral in  Eq. (\ref{eq:b4}) extends only over the regions around $\theta \approx 0$, or $\theta \approx \pi$, and for $r\approx d$.

For the case of an axisymmetric elongational flow defined through Eq. (\ref{eq:s3}) we will consider that the  magnetic field is oriented parallel to the z-axis. In this case, the pair correlation function (\ref{eq:b2}) is given by

\begin{equation}\label{eq:s9}
g(r) = \left\{\begin{array}{cl}
\exp\left\{\left(\frac{P}{2}x^2+\lambda x^{-3}\right)(3\cos^2\theta -1)\right\}&\mbox{; $x\geq 1$}\\
\\
0 &\mbox{; $x<1$}
\end{array}\right.
\end{equation}

\noindent By substitution of Eq. (\ref{eq:s9}) into Eq. (\ref{eq:b4}), we can obtain the asymptotic  expression of $B(T)$

\begin{equation}\label{eq:s10}
B = -\frac{\pi d^3\exp{(2\lambda+P)}}{(6\lambda-2P)(3\lambda+\frac{3}{2}P)}  \; ,
\end{equation}

\noindent where the integration domain of the corresponding integral reduces to the regions around the directions $\theta \approx 0$ or $\theta \approx \pi$, and $r\approx d$.
\section{The average size of a chain}

Our purpose in this section is to ascertain how the size of a chain is modified by the imposed flow. The average size of a chain composed of magnetic grains was first calculated in Ref. $^{\ref{gennes}}$ as a function of the interaction parameter $\lambda$. That approach consists in assuming that the chains of different sizes constitute a mixture of ideal gases in which the different components are identified with the chains having the same number of units. The pressure is then given through its expression for a mixture of ideal gases 

\begin{equation}\label{eq:s11}
p = k_BT\sum_{k=1}^\infty \rho_k\; ,
\end{equation}

\noindent with $\rho_k $ being the concentration of chains containing $k$ dipoles. One defines the average size $N$ of a chain as

\begin{equation}\label{eq:s12}
N \equiv \frac{\sum_{k=1}^\infty k\rho_k}{\sum_{k=1}^\infty \rho_k} = \frac{\rho}{p/k_BT}\; ,
\end{equation}

\noindent where $\rho = \sum_{k=1}^\infty k \rho_k$ is the initial concentration of dipoles. Using in this expression the virial expansion for $p$ up to second order in $\rho$  

\begin{equation}\label{eq:s13}
\frac{p}{k_BT} = \rho (1 + \rho B)\; ,
\end{equation}

\noindent the average size is given by 

\begin{equation}\label{eq:s14}
N = \frac{1}{1 - \rho\vert B(T)\vert}\; ,
\end{equation}

\noindent where we have taken into account the negative character of the second virial coefficient. This approach only provides the correct size of a chain for small values of $\lambda$. At larger values of that parameter $N$ diverges thus indicating that the actual kinetics of the process is not satisfactorily described in that domain. In order to avoid this imbalance between theory and experiment, a different approach, based essentially on the existence of a dynamical equilibrium among different species, was proposed $^{\ref{cebers}}$. The kinetic process for which chains are formed is modeled as a set of coupled binary chemical reactions, in which two chains composed of $k-n$ and $n$ particles may 'react' resulting in a chain of size $k$ or vice versa. After defining the chemical potential of each species or energy per chain, and assuming the condition of chemical equilibrium one achieves a recursion relationship for $\rho_k$ in terms of $\rho_1$, which allows one to express the average size as 

\begin{equation}\label{eq:s15}
N = \frac{2\rho \vert B\vert}{\sqrt{1 + 4\rho\vert B\vert} - 1}\; ,
\end{equation}
\noindent where $\rho= \frac{\phi}{\pi d^3/6}$,with $\phi$ being the initial volume fraction of particles. Both theories report an increase of $N$ as a function of the interaction parameter. The main difference lies in the fact that the second theory predicts the existence of finite chains for values of $\lambda$ larger than the ones corresponding to the first theory, for which N diverges. Divergence of the length of the chain occurs at $\lambda\simeq 2$ in Eq. (\ref{eq:s14}), and at $\lambda\simeq 6$ for the case of Eq. (\ref{eq:s15}). Thus, to analyze the behavior of $N$ as a function of the elongational rate we will use the second approach, exhibiting a wider domain of physical values of $\lambda$.

The expressions of the virial coefficient $B$ obtained for the different situations examined in section 3 can now be used in Eq. (\ref{eq:s15}) to compute the corresponding sizes of the chain. In case a), the averaged size of a chain becomes

\begin{equation}\label{eq:s17}
N = 12\phi \frac{\exp (2\lambda+P/2)}{(6\lambda -P)\sqrt{(3\lambda +P)(3\lambda +P/2)}}\left\{-1 +\sqrt{1 + 24\phi \frac{\exp (2\lambda+P/2)}{(6\lambda -P)\sqrt{(3\lambda +P)(3\lambda +P/2)}}}\right\} ^{-1}\; .
\end{equation}

\noindent This result has been depicted in Fig. 1, from which one observes an enhancement in the length of the chain with respect to the case at zero elongation.

In case b), N is found to be

\begin{equation}\label{eq:s16}
N = \frac{2\phi}{3\lambda^2} \exp (2\lambda+P^2/2\lambda)\left\{-1 + \sqrt{1 +\frac{4\phi}{3\lambda^2} \exp (2\lambda +P^2/2\lambda)}\right\} ^{-1} \; .
\end{equation}

\noindent and has been represented in Fig. 2. This figure clearly shows corrections smaller than the ones observed in the previous case. This fact may be interpreted as a consequence of the loss of symmetry of the system inherent to the orientation of the magnetic field with respect to the flow. From the inset in this figure, we can realize that the effect of the flow is to drag away the particles towards directions different to that of the magnetic field. For this reason a stronger magnetic field, able to balance out the hydrodynamic forces, must be applied, and consequently the effect of the flow is less important than in the previous case. For case c) in which the averaged size of the chain is given by

\begin{equation}\label{eq:s18}
N = \frac{\phi \exp(2\lambda)}{\lambda\sqrt{9\lambda^2- P^2/4}}\left\{-1 +\sqrt{1 + \frac{2\phi \exp(2\lambda)}{\lambda\sqrt{9\lambda^2- P^2/4}}}\right\} ^{-1} \; .
\end{equation}
 
\noindent the flow does not introduce any significant  correction to the result at zero elongational rate. This fact is a consequence of the particular orientation of the magnetic field, perpendicular to the plane of the flow. Due to the high strength of the magnetic field, the chains are oriented in a direction parallel to this field, and the chainig process results unaffected by the flow, irrespective of its nature. Finally, when the flow is axisymmetric we obtain 

\begin{equation}\label{eq:s19}
N = 12\phi \frac{\exp (P+2\lambda)}{(6\lambda -2P)(3P/2+3\lambda)}\left\{-1 +\sqrt{1+ 24\phi \frac{\exp (P+2\lambda)}{(6\lambda -2P)(3P/2+3\lambda)}}\right\} ^{-1} \; .
\end{equation}

\noindent This result has been plotted in Fig. 3 and corresponds to the situation in which the presence of the flow influences to a greatest extent the chaining process. Indeed, for the same conditions as in case a), the correction to the value of $N$ found in the absence of flow practically doubles the one obtained in that case. Now, due to the combined action of both magnetic field and flow, the grains become aligned along the symmetry axis of the system, thus favoring the chaining process. By comparison of the inset of this figure with the corresponding one in Figs. 1 and 2 one can realize that this is the more symmetric situation. 

From the cases analyzed previously one infers a common behavior: the presence of the flow results in an enhancement of the length of the chain when the elongational rate increases. In the case of the axisymmetric flow, when both magnetic field and elongation axis are aligned, the corresponding effects on the grains act in favor of the formation of a chain. For this reason, is in this situation in which larger chains may emerge.

\section{conclusions}

In this paper we have analyzed the influence that the presence of an external flow exerts on the distribution of sizes of chain-like aggregates in magnetic colloids. Due to the potential nature of the flow we have considered,  the potential energy of the resulting structures contains a contribution coming from the flow in addition to the dipolar magnetic energy. This fact has implications in the form of the pair correlation function and consequently in the value of the second virial coefficient. To assess that influence we have implemented typical situations in which the magnetic colloid is subjected to conditions inducing elongational flows. The mean size of the chain can be calculated after computation of the second virial coefficient. To this purpose, we may apply different  theories$^{\ref{gennes},   \ref{cebers}}$ originally proposed for a quiescent liquid. We have adopted the theory introduced in Ref. $^{\ref{cebers}}$ providing results valid for larger values of the $\lambda$ parameter.

The potential character of the flow leads to a Boltzmann-like stationary distribution function for the configuration of bounded pairs of particles$^{\ref{kramers}}$, which allows us to perform an equilibrium-like analysis of the aggregation of particles in the presence of flow. Thus, we can obtain the equation of state that depends on the elongational rate, the parameter  characterizing the flow. An equation of state for a liquid in the presence of flow in which the pressure depends on the shear rate was also obtained from thermodynamic grounds and simulations in Ref. $^{\ref{evans}}$.

Among the cases we have discussed, we note that the most favorable situations for chaining are those in which the magnetic field lies along the directions corresponding to the minima of the elongational potentials. In these cases,  magnetic field and  flow cooperate to create stable chains.

An estimate of the dimensionless parameters appearing in our theory, useful for possible experimental implementations, can be carried out by assuming particles of mass density 1 $g/cm^3$ having a radius ranging from $10^{-6}$ $cm$ to $10^{-5}$ $cm$, suspended in water with viscosity $10^{-2}$ poise. We have also considered the  elongational rate ranging from $10$ $s^{-1}$ to $100$ $s^{-1}$, and the magnetization of the particles of $10$ $Oe$ to $100$ $Oe$. In all cases, the temperature has been assumed to be $300$ $K$. Under this conditions, we can estimate the intervals $5\times 10^{-3}< \lambda <6\times 10^2$, and $2\times 10^{-4}< P <2$. The value of the different quantities corresponding to Figs. 1 - 3 are such that the dimensionless parameters belong to those intervals.

Our main conclusion is that the external flow may introduce significant changes in the kinetics of the process resulting in the formation of chains larger than the corresponding ones obtained in the absence of flow. Our analysis has then revealed that the external flow constitutes an important factor in the formation of chain-like structures in magnetic colloids.

\acknowledgments

This work has been
supported by DGICYT of the Spanish Government under grant PB98-1258, and
also by the INCO-COPERNICUS program of the European Commission under
contract IC15-CT96-0719. We want to thank M.C. Miguel and R. Pastor-Satorras for useful comments.

\begin{center} \appendix{{\bf APPENDIX}} \end{center}

In this appendix we indicate how to compute the virial coefficients obtained in section 3. For instance in case a), we assume that the dipoles forming the dimer are one close to other in such a way that $x$ approximates to $x\simeq 1+\xi$,  with $\xi<<1$. In this situation the more important contribution to the integral (\ref{eq:b4}) comes from to directions: $\theta\sim \pi/2$, $\phi\sim \pi/4$ and $\pi/2$, $5\pi/4$, in which the magnetic energy dominates. Expanding  the expression for the factor $\Phi=-U/kT$ in those directions one obtains  

\begin{equation}
\Phi\simeq 2\lambda +P/2-(3\lambda+P/2)(\theta-\pi/2)^2-(3\lambda+P)(\phi-\pi/4)^2-(6\lambda-P)\xi
\end{equation}

\noindent and 

\begin{equation}
\Phi\simeq 2\lambda +P/2-(3\lambda+P/2)(\theta-\pi/2)^2-(3\lambda+P)(\phi-5\pi/4)^2-(6\lambda-P)\xi\; ,
\end{equation}

\noindent respectively. 

To compute the virial coefficient one rewrites expression (\ref{eq:b2})

\begin{equation}
\label{eq:apen1}
B\simeq -\frac{1}{2}\int_{\Omega} d^3{\bf r}\exp(\Phi) \simeq -\frac{d^3}{2}\exp(2\lambda +P/2)I_1I_2I_3
\end{equation}

\noindent where $I_1$, $I_2$ and $I_3$ are the following integrals

\begin{equation}
\label{eq:apen2}
I_1=\int_0^\infty d\xi(1+2\xi)\exp[-(6\lambda-P)\xi]=\frac{1}{6\lambda-P}\left(1+\frac{2}{6\lambda-P}\right)
\end{equation}

\begin{equation}
\label{eq:apen3}
I_2=\int_0^\pi d\theta\sin\theta\exp[-(3\lambda+P/2)(\theta-\pi/2)^2]\simeq\sqrt{\frac{\pi}{3\lambda+P/2}}\exp[-1/4(3\lambda+P/2)]
\end{equation}

\noindent and 

\begin{equation}
\label{eq:apen4}
I_3=\int_{-\pi/4}^{3\pi/4}d\phi\exp[-(3\lambda+P)(\phi-\pi/4)^2]+\int_{3\pi/4}^{7\pi/4}d\phi\exp[-(3\lambda+P)(\phi-5\pi/4)^2]\simeq 2\sqrt{\frac{\pi}{3\lambda+P}}
\end{equation}

\noindent The expression (\ref{eq:s4}) for the second virial coefficient comes straightforwardly by combining equations (\ref{eq:apen1})-(\ref{eq:apen4}). The  virial coefficients for the remaining cases can be computed in a similar way.

\pagebreak

{\bf FIGURE CAPTIONS}

Figure 1. Averaged size of a chain  as a function of $\lambda$ for  different values of $P$, and for $\phi=0.14$. The direction of the magnetic field form an angle of 45 degrees with the x-axis. The inset shows the corresponding orientation of the chain.\\

Figure 2. Averaged size of a chain as a function of $\lambda$ for different values of $P$, and for $\phi=0.14$. The magnetic field lies parallel to the x-axis. The inset indicates the orientation of the chain.\\

Figure 3. Averaged size of a chain  as a function of $\lambda$ for four different values of $P$, and for $\phi=0.14$. Now the magnetic field is parallel to the axis of the elongational flow. The inset also shows the position of the chain.

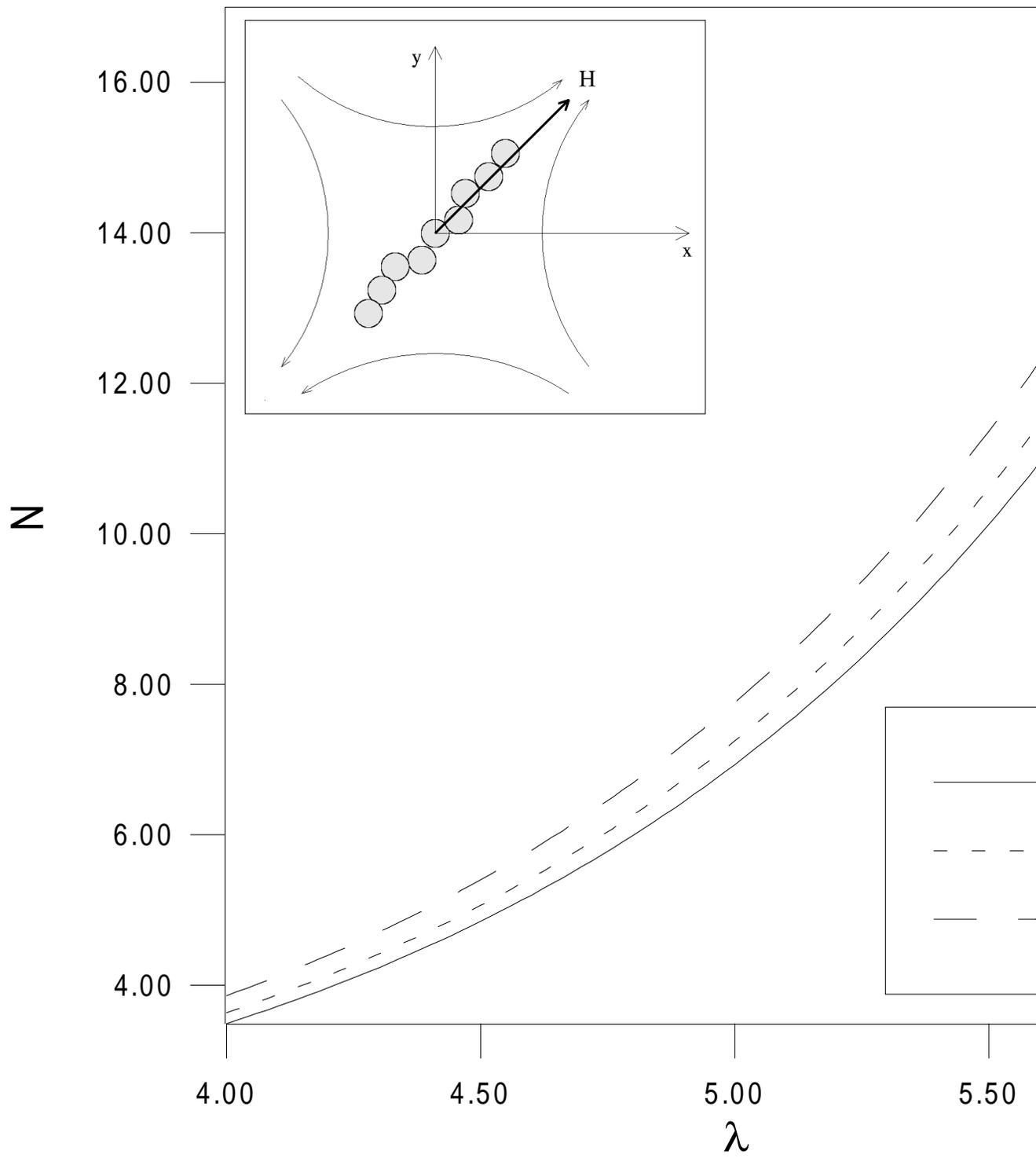
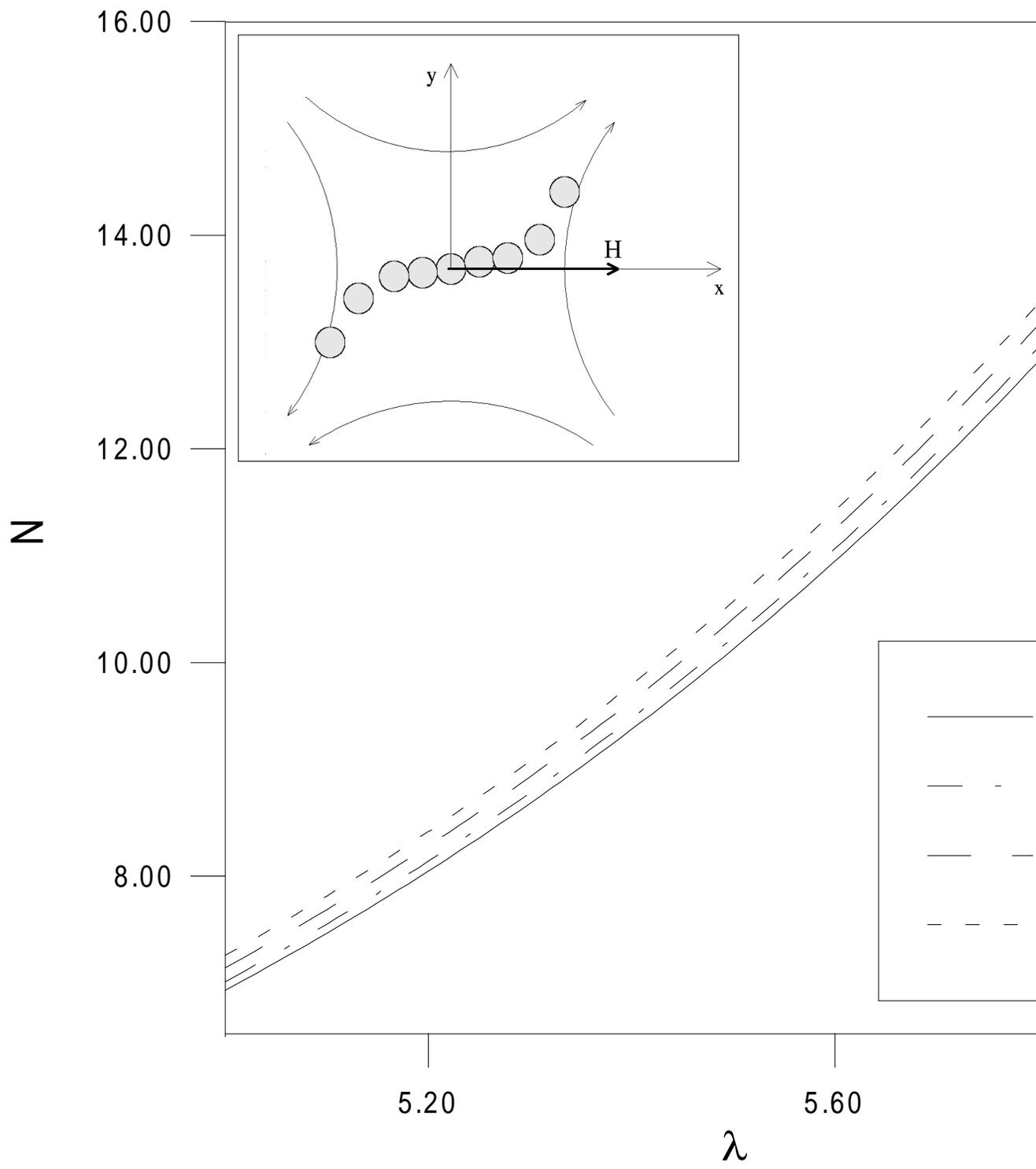
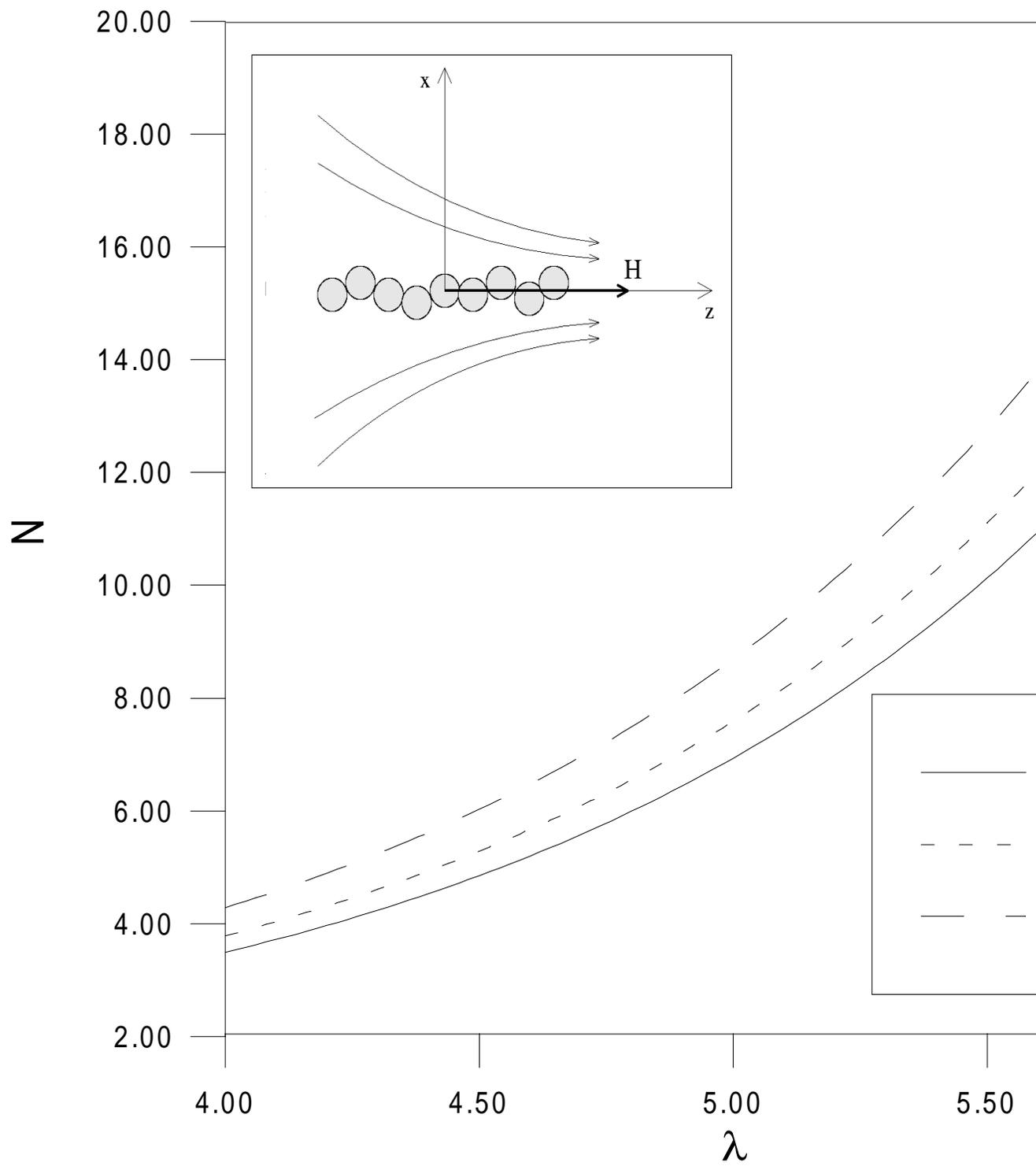
\end{document}